\documentclass{article} 
\usepackage{amsmath,amssymb,pxfonts,graphicx}
\title{Efficient price dynamics in a limit order market: an utility indifference approach\thanks{
This paper is an output of one of the 2012-2013 Project Researches at 
Institute of Economic Research, Kyoto Univerisity as the Joint Usage and Research Center.
The author is grateful to the project members Chiaki Hara, Masaaki Kijima, Katsumasa Nishide, 
Akihisa Tamura and Yuan Tian for their insightful comments and suggestions.
This work was also supported by 
 Japan Society for the Promotion of Science, KAKENHI Grant Number 25245046.}} 
\author{Masaaki Fukasawa\thanks{
M. Fukasawa
\newline
Department of Mathematics, and Center for the Study of Finance and Insurance
\newline
Osaka University
\newline
1-1 Machikaneyama, Toyonaka, Osaka, 560-0043 JAPAN
\newline
Email: fukasawa@math.sci.osaka-u.ac.jp
}} 
\newtheorem{prop}{Proposition}
\newtheorem{lem}{Lemma} \newtheorem{thm}{Theorem}
\begin{document}
\maketitle

\begin{abstract}
 We construct an utility-based dynamic 
asset pricing model for a limit order market.
The price is nonlinear in volume and subject to market impact.
We solve an optimal hedging problem under the market impact and derive the dynamics of the efficient price, that is, the asset price when a representative liquidity demander follows an optimal strategy. 
We show that a Pareto efficient allocation is achieved under a completeness condition.
We give an explicit representation of the efficient price 
for several examples.
In particular, we observe that the volatility of the asset depends on the convexity of an initial endowment.
Further, we observe that 
an asset price crash is invoked by an endowment shock.
We establish a dynamic programming principle under an incomplete framework.
\end{abstract}

 quadratic backward stochastic differential equation; Burgers' equation. \\



\section{Introduction}

We are interested in the dynamics of an asset price in a limit order market.
We aim at constructing a tractable model which captures endogenously such phenomena observed in actual markets as nonlinearity in liquidation, permanent market impact and a flash crash due to illiquidity. The liquidity issue has been pointed out as a major risk which standard models of financial engineering have not taken into consideration. The failures of several financial institutions are often attributed to the uncovered liquidity risk. The liquidity crisis is a rare event; an exogenous statistical modeling of liquidity costs is therefore not sufficient for preparing ourselves for a  future crisis.
An economic consideration is required for a deeper understanding of the liquidity risk.
This paper provides an utility-based asset pricing model with  analytically tractable structure.
\\

In a limit order market, the roles of supplier and demander of liquidity are not symmetric.
A liquidity supplier submits a {\it limit order} that quotes a price for a specified volume of an asset.
They can trade with each other by using limit orders to maximize their own utilities. Once an equilibrium is achieved,
no more trade occurs among them.  However, a liquidity supplier still has  an incentive to submit a limit order 
as long as the corresponding transaction improves her utility.
The remained limit orders form a price curve which is a nonlinear function of volume.
A liquidity demander submits a {\it market order} to buy or sell any amount of the asset according to the price curve.
Taking a Bertrand-type competition among liquidity suppliers into account, it is then reasonable to begin with modeling the price curve as the utility indifference price for a representative liquidity supplier (RLS hereafter).\\

If the RLS is risk-neutral, then the utility indifference price of an asset 
coincides with the expected value of the future cash-flow associated with the asset.
The price curve then becomes linear in volume. This simplest framework was adopted by many studies such as Glosten and Milgrom~\cite{GM}.
In this study,  we assume the RLS to be risk-averse in managing her inventory, which results in nonlinear pricing with market impact.\\

This approach differs from  the classical works including Garman~\cite{Gar1}, Amihud and Mendelson~\cite{AM}, Ho and Stoll~\cite{HS}, Ohara and Oldfield~\cite{OO}, where a price quote is a solution of an utility maximization problem for a market maker with exogenously given order-flow. 
Here, we solve an utility maximization problem for a representative liquidity demander (RLD hereafter). Therefore, an order-flow is endogenously determined.
Our model is closely related to the one considered in Bank and Kramkov~\cite{BK1,BK2},
where they analyzed the market impact of a large trade and formulated a nonlinear stochastic integral as the profit and loss associated with a given strategy of a large trader. Here, we aim at deriving the dynamics of the {\it efficient price}, that is
the asset price when  the RLD follows an optimal strategy.
Unlike standard optimal investment or hedging problems, the asset price is nonlinear in volume and depends on the trading strategy.
If the set of the liquidity demanders consists of price takers and a single large investor, 
the optimal strategy can be interpreted as the solution of an optimal investment or hedging problem for the large investor under a simplified version of the model introduced by
Bank and Kramkov~\cite{BK1,BK2}.
The model represents permanent market impact, while instantaneous or temporary market impact models have been extensively considered in the literature. See e.g. Cetin et al.~\cite{Cetin}, Fukasawa~\cite{F}, Gu\'eant~\cite{Gueant} and the references therein.
In the most of the preceding studies, the structure of market impact was exogenously modeled.
Here, as in Bank and Kramkov~\cite{BK1,BK2}, the market impact is endogenous.\\

We consider a hypothetical market where all traded securities expire at a maturity and yield cash-flow, of which the joint probability distribution is given exogenously. This can be seen as a model of a commodity futures market.
A standard theory shows that a commodity future price is determined by an arbitrage relation to the spot price of the commodity. This arbitrage pricing however does not work for a commodity of which the inventory costs are so expensive, such as oil or electricity.
A major factor which determines the future price of such a commodity would be  the probability distribution of the spot price at the maturity, which is the idea of our model.\\

In Section~2, we give a rigorous formulation of the problem.
In Section~3, we consider a L\'evy-driven market as an example which admits explicit results.
In particular, we observe that an optimal risk allocation between the RLS and the RLD is achieved in a specific situation. We see also that an endowment shock can invoke a price crash with severe illiquidity.
In Section~4, we solve the optimal hedging problem under a condition of completeness,
where the optimal risk allocation is achieved.
In particular, we construct a perfect hedging strategy under the market impact.
We observe that the volatility of the efficient price process depends on the convexity of an initial endowment to be hedged. We see also that a price crash is invoked by a propagated endowment shock.
In Section~5, under an incomplete situation,
we establish a dynamic programming principle in a discrete time framework and give its applications.

\section{Exponential utility indifference pricing}
Here we rigorously formulate the problem.
Let $(\Omega, \mathcal{F}, \mathbb{P}, \{\mathcal{F}_t\}_{t \in [0,1]})$ be a filtered probability space satisfying the usual assumptions
with $\mathcal{F}_0$ composed of the null sets.
The time $1$ stands for the maturity of all securities and we set $\mathcal{F} = \mathcal{F}_1$.
Assume there exists a regular conditional probability measure given $\mathcal{F}_t$ for each $t \in [0,1)$.
By $\mathbb{E}[F| \mathcal{F}_t]$, we always mean the expectation of a random variable $F$ with respect to this regular conditional probability measure.
Let 
\begin{equation*}
S = (S^1,\dots,S^d)^{\mathrm{T}}
\end{equation*}
be an $\mathcal{F}_1$-measurable random vector which stands for 
the cash-flow of $d$ securities at time $1$.
We assume $0$ dividend rate and $0$ risk-free rate.\\

The RLS  evaluates future cash-flow $F$ at time $1$ by
an exponential utility
\begin{equation}\label{Pi}
\Pi_t(F) = -\frac{1}{\gamma} \log \mathbb{E}[\exp(-\gamma F)|
 \mathcal{F}_t]
\end{equation}
at $t\in [0,1]$, where $\gamma \geq 0$ is a parameter of risk-aversion. 
The exponential utility with $\gamma = 0$ is interpreted as the limit of (\ref{Pi}) as $\gamma \to 0$. 
More specifically, under a suitable condition on $F$, $\Pi_t(F) = \mathbb{E}[F|\mathcal{F}_t]$ when $\gamma = 0$, which is the case that the RLS is risk-neutral.  An important property of the general exponential utility is the cash-invariance:
\begin{equation*}
 \Pi_t(F+C) = \Pi_t(F) + C
\end{equation*}
for any $\mathcal{F}_t$-measurable random variable $C$.
Another important property is the quasi-concavity: if
$\Pi_t(F^1) \geq 0$ and $\Pi_t(F^2) \geq 0$, then 
$\Pi_t(\lambda F^1 + (1-\lambda)F^2) \geq 0$ for any $\lambda \in [0,1]$.
Together with the cash-invariance property, the quasi-concavity implies the concavity:
\begin{equation*}
\begin{split}
& \Pi_t(\lambda F^1 + (1-\lambda)F^2)
-\lambda \Pi_t(F^1) - (1-\lambda)\Pi_t(F^2)
\\ &=
\Pi_t(
\lambda (F^1 - \Pi_t(F^1)) +
(1-\lambda) (F^2 - \Pi_t(F^2))
) \geq 0.
\end{split}
\end{equation*}

The  RLS is initially endowed with a risky asset which yields cash-flow at
time $1$, denoted by $G$.
If the RLS is holding $z = (z^1,\dots,z^d)$ units of the $d$ securities as her inventory 
at time $t \in [0,1]$, 
then for a volume  $y = (y^1,\dots,y^d)$,
she quotes a price  according to  the utility indifference principle:
\begin{equation*}
\begin{split}
P_t(z,y) =& \inf\{p \in \mathbb{R}; \Pi_t(G +  zS -yS
 +p) \geq \Pi_t(G + z S) 
\} \\
=& \Pi_t(G +  zS) -
\Pi_t(G +  (z-y)S).
\end{split}
\end{equation*}
Note that the utility indifference price does not depend on the amount
of cash held by the RLS due to that her utility is exponential.
If the RLS is risk-neutral, then $P_t(z,y) = y\mathbb{E}[S|\mathcal{F}_t]$. Otherwise,
the price depends on the inventory $z$ of the securities, 
which describes permanent market impact. For all $t$ and $z$, $P_t(z,y)$ is a convex function of $y$ with $P_t(z,0) = 0$.
This implies in particular that
\begin{equation*}
-P_t(z,-y) \leq P_t(z,y) 
\end{equation*}
for any $y$ and $z$, which means that the selling price for an amount is higher than or equal to the buying price for the same amount.
This represents bid-ask spread that is a measure of market liquidity.
\\

The RLD also evaluates future cash-flow $F$ by an exponential utility
\begin{equation*}
U_t(F) = -\frac{1}{c} \log \mathbb{E}[\exp(-c F)| \mathcal{F}_t]
\end{equation*}
at time $t\in [0,1]$, where $c \geq 0$ is a parameter of risk-aversion.
We allow $c = \infty$ by taking the limit $c \to \infty$.
 The RLD is initially endowed
with a risky asset which yields cash-flow at time $1$, denoted by $H$.
Let $A$ be a subset of $\mathbb{R}^d$ with $0 \in A$. 
We suppose $\Pi_t(G + aS)$ is finite for all $a \in A$ and $t \in [0,1]$.
Let
$\mathcal{S}_A$ be the set of the $d$-dimensional simple predictable processes $Y$
with $Y_{0} = 0$ and $Y_t \in A$ for all $t\in [0,1]$.
The RLD  is allowed to take any element $Y = (Y^1,\dots,Y^d) \in \mathcal{S}_A$ as her trading strategy, where
$Y^i_t$ stands for the number of shares for the $i$-th security 
held by the RLD at time $t$.
The price for a volume $y = (y^1,\dots,y^d)$ of the $d$ securities at time $t$ is
$P_t(-Y_t,y)$. This is because the RLS holds $-Y_t$ units of the
securities due to the preceding trades with the RLD. 
Then the profit and loss at time $1$ associated with $Y \in \mathcal{S}_A$ 
is given by
\begin{equation*}
 \mathcal{I}(Y) = Y_1S - \sum_{0 \leq t < 1} P_t(-Y_t, \Delta Y_t). 
\end{equation*}
The problem of the RLD is to maximize the utility of her terminal wealth
\begin{equation*}
U_0(H + \mathcal{I}(Y))
\end{equation*}
in $Y$ on a suitably extended set of $\mathcal{S}_A$. \\

Naturally arise the following questions:
\begin{enumerate}
\item Does a dynamic programming principle hold for this utility maximization problem ?
\item How is the optimal strategy $Y^\ast$ characterized ?
\item How does the {\it efficient price} $P_t(-Y^\ast_t, y)$ behave ?
\item How does the {\it efficient infinitesimal  price per unit} (EIPU hereafter)
\begin{equation*}
S^\ast = (S^{\ast,1},\dots, S^{\ast, d}), \ \ 
 S^{\ast,i}_t = \frac{\partial}{\partial y^i} \bigg|_{y=0}
P_t(-Y^\ast_t,y)
\end{equation*}
behave ?
\item How does the {\it efficient price convexity}
\begin{equation*}
C^{ij}_t = \frac{\partial^2 }{\partial y^i \partial y^j} \bigg|_{y=0}
P_t(-Y^\ast_t,y), \ \ i, j = 1,\dots, d
\end{equation*}
behave ?
\end{enumerate}
The EIPU is interpreted as the asset price vector
 for small investors who are price takers.
The efficient price convexity is a measure of infinitesimal illiquidity.

\section{L\'evy-driven markets}

Here we give an example which admits explicit computation.
We will see an optimal risk allocation between the RLS and the RLD is achieved in some cases, 
while not in other cases.
 Let  $d = 1$,
$c, \gamma \in (0,\infty)$
and $S = X_1$ for a L\'evy process $X$ with $X_0 = 0$.
The characteristic function of $X_t$ admits the L\'evy-Khintchine representation
\begin{equation*}
\frac{1}{t}\log \mathbb{E}[e^{iuX_t}]=
 iub-\frac{1}{2}\sigma^2 u^2 + \int (e^{iux}-1-iuh(x))\nu(\mathrm{d}x),
\end{equation*}
where $b, \sigma \in \mathbb{R}$, $h(x) = x1_{\{|x|\leq 1\}}$ and $\nu$ is a measure on $\mathbb{R}$ with
\begin{equation*}
 \int 1 \wedge x^2 \nu(\mathrm{d}x) < \infty.
\end{equation*}
Assume that there exists an interval $U \subset \mathbb{R}$ such that
\begin{equation*}
 \int e^{-ux}1_{\{|x| > 1\}}\nu(\mathrm{d}x) < \infty
\end{equation*}
for all $u \in U$. Then, 
\begin{equation*}
 \int |e^{-ux}-1 + uh(x)|\nu(\mathrm{d}x) < \infty
\end{equation*}
for all $u \in U$ and by Theorem~25.3 of Sato~\cite{Sato}, 
\begin{equation}\label{PiLevy}
 \Pi_t(zX_1) = zX_t + \frac{1-t}{\gamma}\kappa(\gamma z),
\end{equation}
for all $z \in U/\gamma$, where
\begin{equation*}
 \kappa(u) = bu - \frac{1}{2}\sigma^2u^2 - \int (e^{-ux}-1 + uh(x))\nu(\mathrm{d}x).
\end{equation*}
Since $\Pi_0$ is a concave functional, $\kappa$ is a concave function with $\kappa(0)=0$.
For example,
\begin{equation*}
 \kappa(u) = bu - \frac{1}{2}\sigma^2u^2, \ \ U= \mathbb{R}
\end{equation*}
if $X_t = bt + \sigma W_t$, where $W$ is a standard Brownian motion. We have
\begin{equation*}
 \kappa(u) = \beta \log(1 + u/\alpha), \ \ U = (-\alpha,\infty)
\end{equation*}
if $X$ is a gamma process with $X_1$ following the gamma distribution with rate $\alpha>0$ and shape $\beta > 0$. Another example is
\begin{equation*}
\kappa(u) = r u^\alpha, \ \ U = [0,\infty)
\end{equation*}
when $X$ is a one-sided stable process with rate $r>0$ and exponent $\alpha \in (0,1)$.\\

Suppose $G=aS$ for $a \in U/\gamma$ and $A \subset a - U/\gamma$. 
Then, by (\ref{PiLevy}),
\begin{equation*}
 P_t(z,y) = yX_t + \frac{1-t}{\gamma}(\kappa(\gamma(a+z))-\kappa(\gamma(a+z-y))),
\end{equation*}
for $z \in -A$ and $y \in A +z$ and so, 
\begin{equation*}
\begin{split}
 P_t(-Y_t,\Delta Y_t) =& X_t\Delta Y_t - \frac{1-t}{\gamma}\Delta\kappa(\gamma(a-Y_t)) \\
 =& X_{t-}\Delta Y_t  + \Delta X_t \Delta Y_t - \frac{1-t}{\gamma}\Delta\kappa(\gamma(a-Y_t))
\end{split}
\end{equation*}
for $Y \in \mathcal{S}_A$.
By integration by parts,
\begin{equation}\label{PL}
 Y_1X_1 - \sum_{0 \leq t < 1} P_t(-Y_t,\Delta Y_t)
=   \int_0^1 Y_t \mathrm{d}X_t + \frac{1}{\gamma}\int_0^1 (\kappa(\gamma(a-Y_t))-\kappa(\gamma a))\mathrm{d}t.
\end{equation}

Note that if $0$ is an interior point of $U$, then $\kappa$ is differentiable at $0$ and
by letting $\gamma \to 0$, 
\begin{equation*}
 P_t(z,y) \to y\tilde{X}_t, \ \ \tilde{X}_t = X_t + (1-t)\mathbb{E}[X_1].
\end{equation*}
The process $\tilde{X}$ is a martingale and  it was the asset price process when the RLS was risk-neutral.
The right hand side of (\ref{PL}) can be written as
\begin{equation}\label{PL2}
 \int_0^1 Y_t \mathrm{d}\tilde{X}_t 
+\int_0^1 Y_t \left\{\kappa^\prime(0) -\frac{\kappa(\gamma a) -\kappa(\gamma (a-Y_t))}{\gamma Y_t} \right\}\mathrm{d}t.
\end{equation}
The first term is of the familiar form of profit and loss associated with the trading strategy $Y$
when $\tilde{X}$ was the price process.
The second term can be both positive and negative and converges to $0$ as $\gamma \to 0$. 
\\

Now, consider  $H$ of the form
\begin{equation*}
 H = h + \int_0^1 H^\prime_t\mathrm{d}X_t,
\end{equation*}
where $h \in \mathbb{R}$ and $H^\prime$ is a simple predictable
bounded process independent of $X$. 
This specification is just to obtain an explicit result.
If one interprets $\mathrm{d}X_t$ as an economic signal at time $t$, then $H^\prime_t$ determines its contribution to the initial endowment $H$.\\

If $c(H^\prime+Y)$ takes values in $U$, then
the terminal wealth of the RLD is
\begin{equation*}
 \begin{split}
& h + \int_0^1(H_t^\prime + Y_t)\mathrm{d}X_t + \frac{1}{\gamma}\int_0^1 (\kappa(\gamma(a-Y_t))-\kappa(\gamma a))\mathrm{d}t
\\ = &  \ 
h + \int_0^1(H_t^\prime + Y_t)\mathrm{d}X_t - \frac{1}{c}\int_0^1 \kappa(c(H_t^\prime + Y_t))\mathrm{d}t 
\\ &+ \frac{1}{c}\int_0^1 \kappa(c(H_t^\prime + Y_t))\mathrm{d}t +
\frac{1}{\gamma}\int_0^1 (\kappa(\gamma(a-Y_t))-\kappa(\gamma a))\mathrm{d}t\\
\leq & \ h + \int_0^1(H_t^\prime +Y_t)\mathrm{d}X_t - \frac{1}{c}\int_0^1 \kappa(c(H_t^\prime + Y_t))\mathrm{d}t 
\\
& \ + \frac{c+\gamma}{c\gamma}\int_0^1 \kappa\left(\frac{c\gamma}{c+\gamma}(a + H_t^\prime)\right)\mathrm{dt} - \frac{1}{\gamma}\kappa(\gamma a).
 \end{split}
\end{equation*} 
The last inequality follows from the convexity of $\kappa$ and the equality is attained when
\begin{equation}\label{opt}
 Y_t = Y^\ast_t :=\frac{\gamma}{c + \gamma} a - \frac{c}{c+\gamma}H_t^\prime.
\end{equation}
Since $H^\prime$ is independent of $X$, we have
\begin{equation*}
 \mathbb{E}\left[
\exp\left\{
-c \int_0^1 (H_t^\prime + Y_t)\mathrm{d}X_t + \int_0^1 \kappa(c(H_t^\prime + Y_t))\mathrm{d}t
\right\} \bigg| H^\prime
\right] = 1
\end{equation*}
if, say, $Y$ is bounded. See Kallsen and Shiryaev~\cite{KS}.
If $A$ is bounded but enough large for $\mathcal{S}_A$ to include $Y^\ast$ given by (\ref{opt}),
then
\begin{equation*}
\begin{split}
\max_{Y \in \mathcal{S}_A}U_0(H + \mathcal{I}(Y)) =&
U_0(H + \mathcal{I}(Y^\ast))  \\ =&
 U_0\left(h +
\frac{c+\gamma}{c\gamma}
\int_0^1 \kappa\left(\frac{c\gamma}{c+\gamma}(a + H^\prime_t)\right)\mathrm{d}t
\right) -\frac{1}{\gamma}\kappa(\gamma a).
\end{split}
\end{equation*}
To interpret this identity, let us introduce
\begin{equation*}
U^\ast_{H^\prime}(F) = - \frac{c + \gamma}{c\gamma} \log \mathbb{E}\left[
\exp\left\{-
\frac{c\gamma}{c+\gamma}F
\right\} \bigg| H^\prime
\right]
\end{equation*}
for random variable $F$. Then,
\begin{equation*}
U^\ast_{H^\prime}(G+H) = h + \frac{c+ \gamma}{c\gamma}\int_0^1 \kappa\left(\frac{c\gamma}{c+\gamma}(a+H^\prime_t)\right)\mathrm{d}t
\end{equation*}
and so,
\begin{equation}\label{alloc}
\max_{Y \in \mathcal{S}_A}U_0(H + \mathcal{I}(Y)) 
=
U_0(U^\ast_{H^\prime}(G+H))-\Pi_0(G).
\end{equation}
Note that
\begin{equation*}
\frac{c\gamma}{c+ \gamma} <  c, \ \ 
U_0(U^\ast_{H^\prime}(G+H))  \leq
U^\ast_0(U^\ast_{H^\prime}(G+H))  =
U^\ast_0(G+H),
\end{equation*}
where
\begin{equation}\label{uast}
U^\ast_t(F) = - \frac{c + \gamma}{c\gamma} \log \mathbb{E}\left[
\exp\left\{-
\frac{c\gamma}{c+\gamma}F
\right\} \bigg|\mathcal{F}_t
\right]
\end{equation}
for random variable $F$.
This functional $U^\ast$ is known as the aggregated utility 
under optimal risk allocation (see e.g., Barrieu and El Karoui~\cite{BE}).
The above computation shows that the optimal risk allocation between the RLS and the RLD 
is achieved by a dynamic trading strategy $Y^\ast$
if $H^\prime$ is deterministic, while some utility
can be lost otherwise. \\

Since the optimal strategy $Y^\ast$ of the RLD is given by (\ref{opt}),
the efficient price and the EIPU are respectively
\begin{equation*}
\begin{split}
&P_t(-Y^\ast_t,y) = yX_t + \frac{1-t}\gamma
\left(
\kappa\left(
\frac{c\gamma}{c+\gamma}(a+H_t^\prime)
\right)
-\kappa\left(
\frac{c\gamma}{c+\gamma}(a+H_t^\prime)-\gamma y
\right)
\right),\\
 &S_t^\ast =
  \frac{\partial}{\partial y} \bigg |_{y=0} P_t(-Y^\ast_t,y) = 
 X_t + (1-t) \kappa^\prime\left(
\frac{c\gamma}{c+\gamma}(a+H_t^\prime)
\right)
\end{split}
\end{equation*}
if $\kappa$ is differentiable. The latter can rewritten as
\begin{equation*}
 S^\ast_t = \tilde{X}_t - (1-t)
\left(\kappa^\prime(0) - \kappa^\prime\left(
\frac{c\gamma}{c+\gamma}(a+H_t^\prime)\right) 
\right).
\end{equation*}
Since the first term is a martingale, the second term is interpreted as
the {\it risk premium}.
On intervals where $H^\prime$ is constant, we have
\begin{equation*}
\mathrm{d}S^\ast_t =  
\left(\kappa^\prime(0) - \kappa^\prime\left(
\frac{c\gamma}{c+\gamma}(a+H_t^\prime)\right) 
\right)\mathrm{d}t + \mathrm{d}\tilde{X}_t.
\end{equation*}
Since $\kappa$ is concave, the risk premium is nonnegative if the aggregated initial endowment $a + H^\prime_t$ is nonnegative. 
A positive (resp. negative) endowment shock $\Delta H^\prime$ induces a nonpositive (resp. nonnegative) jump of $S^\ast$ and a nonnegative (resp. nonpositive) jump of the risk premium.
The efficient price convexity is computed as
\begin{equation*}
  \frac{\partial^2}{\partial y^2} \bigg |_{y=0} P_t(-Y^\ast_t,y) = -\gamma (1-t)
\kappa^{\prime \prime}\left(
\frac{c\gamma}{c+\gamma}(a+H_t^\prime)
\right) \geq 0
\end{equation*}
if $\kappa$ is twice differentiable. This implies that
the illiquidity level strongly depends on the convexity of $\kappa$ and so, the distribution tail of $S$.
If $k^{\prime \prime}$ is one more times differentiable and the L\'evy measure $\nu$ is spectrally negative,
then $k^{\prime \prime \prime} < 0$ and the skewness of $S$ is negative. 
In this case, since $-\kappa^{\prime \prime}$ is increasing, a positive endowment shock $\Delta H^\prime > 0$ induces
a positive jump of the infinitesimal illiquidity, in other words,
a wider bid-ask spread,  in addition to a drop of $S^\ast$.
\\


\section{Complete markets}
\subsection{The completeness condition}
Here, we assume 
the filtration $\{\mathcal{F}_t\}$ is generated by a $k$-dimensional standard Brownian motion
$W$.  We solve the problem of the RLD under a completeness condition specified below.
We set $A = \mathbb{R}^d$ for brevity. Let $\mathcal{L}$ be the set of random variables $F$ such that
for all $a>0$, $\mathbb{E}[\exp(a|F|)]<\infty$.
Note that $\mathcal{L}$ is a linear space.
We assume that $G, H$ and $S$ are elements of $\mathcal{L}$.
Let $\Pi^y$ be a continuous modification of $\Pi (G-yS)$ and
\begin{equation*}
 Z^y_t = (Z^{y,1}_t,\dots,Z^{y,k}_t), \ \ 
Z^{y,i}_t = - \frac{\mathrm{d}}{\mathrm{d}t} \langle  \Pi^y, W^i \rangle_t
\end{equation*}
for $y \in \mathbb{R}^d$.
The key assumption of the section is the existence of a map
\begin{equation*}
Z: \Omega \times [0,1] \times \mathbb{R}^d \to \mathbb{R}^k
\end{equation*}
with the following properties:
\begin{enumerate}
 \item For each $t \in [0,1]$, the restriction
\begin{equation*}
 Z1_{[0,t]} : \Omega \times [0,t] \times \mathbb{R}^d \to \mathbb{R}^k
\end{equation*}
is $\mathcal{F}_t \otimes \mathcal{B}([0,t]) \times \mathcal{B}(\mathbb{R}^d)$-measurable.
\item There exists $\Omega_0 \in \mathcal{F}$ with $\mathbb{P}(\Omega_0) = 1$ such that 
\begin{equation*}
Z(\omega,t,y) = Z^y_t(\omega)
\end{equation*}
for all $(\omega,t,y) \in \Omega_0 \times [0,1] \times \mathbb{R}^d$.

\item There exists $\Omega_0 \in \mathcal{F}$ with $\mathbb{P}(\Omega_0) = 1$ such that 
$Z(\omega, t, y)$ is continuous in $y$
for each $(\omega, t) \in \Omega_0 \times [0,1]$. 

\item There exists a map
\begin{equation*}
Y^\dagger: \Omega \times [0,1] \times  \mathbb{R}^k \to \mathbb{R}^d
\end{equation*}
such that the restriction 
\begin{equation*}
Y^\dagger |_{[0,t]} : \Omega \times [0,t] \times \mathbb{R}^k \to \mathbb{R}^d
\end{equation*}
 is $\mathcal{F}_t \otimes \mathcal{B}([0,t]) \otimes \mathcal{B}(\mathbb{R}^k)$-measurable for 
each $t \in [0,1]$ and
there exists $\Omega_0 \in \mathcal{F}$ with $\mathbb{P}(\Omega_0) = 1$ such that  
\begin{equation*}
 Z(\omega, t, Y^\dagger(\omega,t,z)) = z
\end{equation*}
for each $(\omega, t,z) \in \Omega_0 \times [0,1] \times \mathbb{R}^k$.
\end{enumerate}
We call this assumption the {\it completeness condition} hereafter. \\

The terminology is justified by considering the risk-neutral case;
when $\gamma = 0$ and $G$ and $S$ are square-integrable, then
\begin{equation*}
 \Pi^y_t = \mathbb{E}[G|\mathcal{F}_t] + y \mathbb{E}[S|\mathcal{F}_t]
= \int_0^t (G^\prime_t + y S^\prime_t)\mathrm{d}W_t
\end{equation*}
almost surely by the It$\hat{\text{o}}$ representation theorem, where $G^\prime$ and $S^\prime$ are progressively measurable processes which are $\mathbb{R}^k$-valued and 
$\mathbb{R}^d \otimes \mathbb{R}^k$-valued respectively. 
In this case, the asset price is linear in volume and
\begin{equation*}
 \mathrm{d}S^\ast_t = S^\prime_t \mathrm{d}W_t.
\end{equation*}
Since
\begin{equation*}
 Z^y_t = - G^\prime - yS^\prime_t,
\end{equation*}
the completeness condition is satisfied if $S^\prime_t$ has rank $k$ as a $d\times k$ matrix and admits a left inverse matrix which is progressively measurable as a process.\\

Another elementary example is the case $G = a + b W_1$ and $S = \alpha + \beta W_1$ 
with $a \in \mathbb{R}$, $b\in \mathbb{R}^k$, $\alpha \in \mathbb{R}^d$ and 
$\beta \in \mathbb{R}^d \otimes \mathbb{R}^k$. Since
\begin{equation*}
 \Pi_t(G-yS) =
a - y\alpha + (b-y\beta)W_t -
\frac{\gamma}{2}(1-t)|b-y\beta|^2,
\end{equation*}
the completeness condition is satisfied if and only if $\beta$ has rank $k$.\\

\begin{lem}\label{ext}
For $Y \in \mathcal{S}_A$, we have 
\begin{equation*}
 \mathcal{I}(Y) = G - \Pi_0(G) - \int_0^1 \frac{\gamma}{2}|Z^Y_t|^2 \mathrm{d}t
+ \int_0^1 Z^{Y}_t \mathrm{d}W_t,
\end{equation*}
where $Z^Y_t(\omega) = Z(\omega,t,Y_t(\omega))$.
\end{lem}
{\it Proof: }
By the completeness condition, we identify $Z^y_t(\omega) = Z(\omega ,t, y)$.
Let $Y \in \mathcal{S}_A$. Denote by
\begin{equation*}
 0 \leq \tau_1 < \tau_2 < \cdots 
\end{equation*}
the stopping times when $Y$ jumps. 
We have $\tau_n = 1$ for sufficiently large $n$ almost surely.
Remember that $Y_{\tau_{j+1}}$ is $\mathcal{F}_{\tau_j}$-measurable.
Note that 
\begin{equation*}
\begin{split}
 \mathcal{I}(Y) =& \ G - \Pi_1(G-Y_1S) - \sum_{0 \leq t < 1} (\Pi_t(G-Y_tS) - \Pi_t(G-Y_{t+}S))\\
=& \ G - \Pi^0_{\tau_1}(G)- 
\sum_{j=1}^{\infty} (\Pi_{\tau_{j+1}}^{(j)} - \Pi_{\tau_j}^{(j)}),
\end{split}
\end{equation*}
where $\Pi^{(j)} = \Pi^y$ with $y = Y_{\tau_{j+1}}$.
As is well-known, $(\Pi^y,Z^y)$ 
is the unique solution of the backward stochastic differential equation (BSDE hereafter)
\begin{equation*}
G -yS = \Pi^y_t + \int_t^1 \frac{\gamma}{2}|Z^y_s|^2 \mathrm{d}s - \int_t^1 Z^y_s \mathrm{d}W_s.
\end{equation*}
To see this, it is enough to observe that
$M = \exp\{-\gamma \Pi^y\}$ is a local martingale with $M_1 = \exp\{-\gamma(G-yS)\}$ and then,
apply the It$\hat{\text{o}}$ representation theorem.
This BSDE representation implies that
\begin{equation*}
 \Pi^y_t - \Pi^y_s = \int_s^t \frac{\gamma}{2}|Z^y_u|^2\mathrm{d}u
-\int_s^t Z^y_u\mathrm{d}W_u
\end{equation*}
and so, the result follows. \hfill//// \\

By Lemma~\ref{ext}, we can extend the definition of $\mathcal{I}(Y)$ onto the set $\mathcal{S}$ of the progressively measurable processes $Y$ with 
\begin{equation*}
 \int_0^1 |Z^Y_t|^2 \mathrm{d}t < \infty
\end{equation*}
almost surely and
\begin{equation*}
 - \int_0^1 \frac{\gamma}{2}|Z^Y_t|^2 \mathrm{d}t
+ \int_0^1 Z^{Y}_t \mathrm{d}W_t \in \mathcal{L}.
\end{equation*}
 The problem is then to find an optimal strategy $Y \in \mathcal{S}$.
\\

\subsection{The hedging strategy for the RLD with $c = \infty$}
\begin{lem}\label{ph}
For any $H \in \mathcal{L}$, there exists  $Y \in \mathcal{S}$ such that
\begin{equation*}
 -H = \Pi_0(G) - \Pi_0(G+H) + \mathcal{I}(Y).
\end{equation*} 
\end{lem}
{\it Proof: }
This is a perfect hedging problem and
amounts to finding the solution $(V, Z^\ast)$ 
of the BSDE
\begin{equation*}
 G + H = V_t + \int_t^1 \frac{\gamma}{2}  |Z^\ast_s|^2\mathrm{d}s
- \int_t^1 Z^\ast_s\mathrm{d}W_s.
\end{equation*}
As before, this quadratic BSDE admits an explicit solution given by
\begin{equation*}
 V_t = \Pi_t(G+H), \ \ 
Z^\ast = (Z^{\ast,1},\dots Z^{\ast,k}), \ \ 
Z^{\ast,i}_t = - \frac{\mathrm{d}}{\mathrm{d}t} \langle V, W^i \rangle_t.
\end{equation*}
By the completeness condition, we can define
a progressively measurable process $Y^\ast$ by
\begin{equation}\label{yast}
 Y^\ast_t(\omega) = Y^\dagger(\omega, t, Z^\ast_t(\omega))
\end{equation}
to have
\begin{equation*}
 Z^{Y^\ast} = Z^\ast.
\end{equation*}
\hfill////\\

By Lemma~\ref{ph}, any claim $-H$ is perfectly replicated by a strategy $Y \in \mathcal{S}$ under the completeness condition.
The {\it replication price} of $-H$ is given by
\begin{equation}\label{rp}
 p(-H) :=  \Pi_0(G) - \Pi_0(G+H).
\end{equation}
This is convex in $-H$, which
can be interpreted as a diversification of the liquidity risk. \\

\begin{thm}\label{thm:ph}
If $c = \infty$, then
\begin{equation*}
\max_{Y \in \mathcal{S}}U_0(H + \mathcal{I}(Y)) = 
U_0(H + \mathcal{I}(Y^\ast)) =  \Pi_0(G+H)-\Pi_0(G),
\end{equation*} 
where $Y^\ast$ is given by (\ref{yast}).
\end{thm}
{\it Proof: }
Recall a well-known representation
\begin{equation*}
 U_0(F) = \inf_{Q \sim \mathbb{P}} \left\{
\mathbb{E}[F \frac{\mathrm{d}Q}{\mathrm{d}\mathbb{P}}]
+ \frac{1}{c} \mathbb{E}[\frac{\mathrm{d}Q}{\mathrm{d}\mathbb{P}}\log \frac{\mathrm{d}Q}{\mathrm{d}\mathbb{P}}]
\right\}
\end{equation*}
for $c \in (0,\infty)$.
Take the limit $c \to \infty$, to have
\begin{equation*}
 U_0(F) = \inf_{Q \sim \mathbb{P}}
\mathbb{E}[F \frac{\mathrm{d}Q}{\mathrm{d}\mathbb{P}}] =  \mathrm{ess.inf} \ F
\end{equation*}
for any bounded random variable $F$ (See e.g., Delbaen~\cite{Del}). 
This means
that the RLD is extremely risk-averse and tries to offset her initial endowment $H$ almost surely.
By Lemma~\ref{ph}, 
\begin{equation*}
 H + \mathcal{I}(Y^\ast) = \Pi_0(G+H)-\Pi_0(G)
\end{equation*}
is constant and so,
\begin{equation*}
\sup_{Y \in \mathcal{S}}U_0(H + \mathcal{I}(Y)) \geq
U_0(H + \mathcal{I}(Y^\ast)) =  \Pi_0(G+H)-\Pi_0(G).
\end{equation*} 
Suppose that there exists $ Y \in \mathcal{S}$ such that
\begin{equation}\label{deny}
\mathrm{ess.inf }( H + \mathcal{I}(Y)) > \Pi_0(G+H)-\Pi_0(G).
\end{equation}
Since
\begin{equation*}
G - \mathcal{I}(Y) = \Pi_0(G)+
\int_0^1 \frac{\gamma}{2}|Z^Y_t|^2 \mathrm{d}t
- \int_0^1 Z^{Y}_t \mathrm{d}W_t,
\end{equation*}
we have
\begin{equation}\label{pig}
 \Pi_0(G-\mathcal{I}(Y)) = \Pi_0(G)
\end{equation}
by the uniqueness of the solution of the BSDE.
From this and (\ref{deny}), we deduce a contradiction as follows:
\begin{equation*}
 \Pi_0(G) = \Pi_0(G-\mathcal{I}(Y)) < \Pi_0(G+H) - (\Pi_0(G+H)-\Pi_0(G)) = \Pi_0(G).
\end{equation*}
 \hfill////

\subsection{The optimal strategy for the RLD with $c < \infty$}
Here we extend Theorem~\ref{thm:ph} to the case $c < \infty$.
First we state a version of the result referred as Borch's theorem in Barrieu and El Karoui~\cite{BE}.
\begin{lem}\label{Borch}
If $c < \infty$, then for any $G, H \in \mathcal{L}$,
 \begin{equation*}
  \sup_{F \in \mathcal{L}}
\{ U_0(F) + \Pi_0(G+H-F)\}
= U_0(F^\ast) + \Pi_0(G+H-F^\ast) = U^\ast_0(G+H),
 \end{equation*}
where
\begin{equation*}
 F^\ast = \frac{\gamma}{c+\gamma}(G+H)
\end{equation*}
and $U^\ast$ is defined by (\ref{uast}).
\end{lem}
{\it Proof: }
Since
\begin{equation*}
 F \mapsto U_0(F) + \Pi_0(G+H-F)
\end{equation*}
is concave, it suffices to observe that
\begin{equation*}
 \frac{\mathrm{d}}{\mathrm{d}\epsilon} \bigg|_{\epsilon = 0}
\{
U_0(F^\ast +\epsilon X) + \Pi_0(G+H - F^\ast - \epsilon X) \}= 0
\end{equation*}
for all $X \in \mathcal{L}$. \hfill////\\

\begin{thm}\label{thm:opt}
If $c < \infty$,
 \begin{equation*}
\max_{Y \in \mathcal{S}}U_0(H + \mathcal{I}(Y)) = 
U_0(H + \mathcal{I}(Y^\ast)) =  U_0^\ast(G+H)-\Pi_0(G),
\end{equation*} 
where $Y^\ast$ is given by
\begin{equation*}
  Y^\ast_t(\omega) = Y^\dagger(\omega, t, \frac{c}{c+\gamma} Z^\dagger_t(\omega)), \ \ 
Z^{\dagger_t,i}_t = - \frac{\mathrm{d}}{\mathrm{d}t} \langle U^\dagger, W^i \rangle_t, \ \ i = 1,\dots,d
\end{equation*}
and $U^\dagger$ is the continuous modification of $U^\ast(G+H)$.
\end{thm}
{\it Proof: }
By definition, 
\begin{equation*}
 H + \mathcal{I}(Y^\ast)
= G+H- \Pi_0(G) + \frac{c}{c + \gamma} \left\{
-\int_0^1\frac{1}{2} \frac{c\gamma}{c+\gamma} |Z^\dagger_t|^2 \mathrm{d}t
 + \int_0^1 Z^\dagger_t \mathrm{d}W_t
\right\}.
\end{equation*}
Since $(U^\dagger, Z^\dagger)$ is the solution of the BSDE
\begin{equation*}
 G+H = U^\dagger_t + \int_t^1 
\frac{1}{2} \frac{c\gamma}{c+\gamma} |Z^\dagger_s|^2 \mathrm{d}s
 - \int_t^1 Z^\dagger_s \mathrm{d}W_s,
\end{equation*}
we have
\begin{equation*}
 H + \mathcal{I}(Y^\ast)
= \frac{\gamma}{c+\gamma}(G+H) + \frac{c}{c+\gamma}U^\ast_0(G+H)-\Pi_0(G).
\end{equation*}
It follows then,
\begin{equation*}
\begin{split}
 U_0(H + \mathcal{I}(Y^\ast)) =& \frac{\gamma}{c+\gamma}U^\ast_0(G+H)
+ \frac{c}{c+\gamma}U^\ast_0(G+H)-\Pi_0(G) \\ =& 
U^\ast_0(G+H) - \Pi_0(G).
\end{split}
\end{equation*}
It remains to show that
\begin{equation*}
 \sup_{Y \in \mathcal{S}}U_0(H + \mathcal{I}(Y)) \leq
U_0(H + \mathcal{I}(Y^\ast)).
\end{equation*}
As before, we have (\ref{pig}) for all $Y \in \mathcal{S}$.
Therefore,
\begin{equation*}
 U_0(H+\mathcal{I}(Y))
\leq U^\ast_0(G+H) - \Pi_0(G+H-(H+\mathcal{I}(Y)))
= U^\ast_0(G+H) - \Pi_0(G)
\end{equation*}
by Lemma~\ref{Borch}. \hfill////\\

Since $c/(c+\gamma) \to 1$ and so, 
$U^\ast_0(G+H) \to \Pi_0(G+H)$ as $c \to \infty$, Theorem~\ref{thm:opt} extends Theorem~\ref{thm:ph}.
As mentioned in the previous section, $U^\ast_0(G+H)$ is the aggregated utility of the aggregated initial endowment.
The above result implies that an optimal risk allocation (Pareto allocation) is achieved by the dynamic trading strategy $Y^\ast$ for general $G, H \in \mathcal{L}$ under the completeness condition.

\subsection{Markov models}
Here we assume in addition that
\begin{equation*}
 S = s(W_1), \ \ G = g(W_1), \ \ H = h(W_1)
\end{equation*}
with Borel functions $s:\mathbb{R}^k \to \mathbb{R}^d$ and $g, h :\mathbb{R}^k \to \mathbb{R}$
to extract a more tractable structure of the optimal strategy.
Define functions $v: [0,1] \times \mathbb{R}^k \to \mathbb{R}$ and
$p : [0,1] \times \mathbb{R}^k \times \mathbb{R}^d \to \mathbb{R}$
as
\begin{equation}\label{KPZ}
\begin{split}
&v(t,w) = - \frac{c+\gamma}{c\gamma} \log
\mathbb{E}\left[\exp\left\{-\frac{c\gamma}{c+\gamma} (G+H) \right\} \big|W_t=w\right],\\
 &p(t,w,y) = 
- \frac{1}{\gamma} \log
\mathbb{E}\left[\exp\left\{-\gamma(G-yS) \right\} \big|W_t=w\right].
\end{split}
\end{equation}
By the assumption on the filtration, we have
\begin{equation}\label{markov}
 v(t,W_t) = \Pi^\ast_t(G+H), \ \ p(t,W_t,y) = \Pi_t(G-yS).
\end{equation}
As is well-known and easily checked, 
$v$ and $p$ are the solutions of the partial differential equations (PDE hereafter)
\begin{equation*}
\begin{split}
 \frac{\partial v}{\partial t} +
 \frac{1}{2}\biggl\{
\Delta v - \frac{c\gamma}{c + \gamma}|\nabla v|^2
\biggr\} = 0, \ \ v(1,w) = g(w) + h(w),\\
 \frac{\partial p}{\partial t} +
 \frac{1}{2}\biggl\{
\Delta p - \gamma |\nabla p|^2
\biggr\} = 0, \ \ p(1,w,y) = g(w) -ys(w),
\end{split}
\end{equation*}
where
\begin{equation*}
 \Delta = \sum_{i=1}^k \frac{\partial^2}{\partial w_i^2}, \ \ 
\nabla = \left(
\frac{\partial}{\partial w_1}, \dots,
\frac{\partial}{\partial w_k}
\right).
\end{equation*}
The completeness condition can be restated as follows:
there exists a measurable function $y^\dagger : \mathbb{R}^k \times [0,1]\times \mathbb{R}^k \to \mathbb{R}^d$ such that
\begin{equation}\label{complete:markov}
 - \nabla p(t,w,y^\dagger(w, t,z)) = z
\end{equation}
for each $(z,t,w) \in \mathbb{R}^k \times [0,1] \times \mathbb{R}^k$.
The optimal strategy given by Theorems~\ref{thm:ph} and \ref{thm:opt} is 
\begin{equation*}
 Y^\ast_t = y^\dagger(W_t,t, - \frac{c}{c+\gamma} \nabla v(t,W_t)).
\end{equation*}
The efficient price, the efficient price per unit and the efficient price convexity are given by
\begin{equation*}
 \begin{split}
&  P_t(-Y^\ast_t,y) = p(t,W_t, Y^\ast_t) -p(t,W_t, Y^\ast_t+y), \\
&  S^\ast_t = - \frac{\partial p }{\partial y^i} (t,W_t,Y^\ast_t), \ \ i = 1,\dots,d, \\
& C_t^{ij} = - \frac{\partial^2 p }{\partial y^i \partial y^j} (t,W_t,Y^\ast_t), \ \ i, j = 1,\dots,d.
 \end{split}
\end{equation*}

\subsection{The volatility of the EIPU}
Here we study through a simple example 
how the volatility of the EIPU is determined by
model parameters.
Let $k=d=1$, $G = gS$, $S = \mu + \sigma W_1$ and
\begin{equation*}
H = a W_1 + \frac{b}{2} W_1^2
= a\frac{S-\mu}{\sigma} + \frac{b}{2}\left(\frac{S-\mu}{\sigma}\right)^2
\end{equation*}
for constants $g, \mu, \sigma, h, a, b \in \mathbb{R}$.
We assume $c + \gamma + c\gamma b > 0$ and $\sigma \neq 0$.
Although $H$ does not belong to $\mathcal{L}$, it is not difficult to see the preceding results are extended to this specific model.
In this case,  $S$ follows a normal distribution and
\begin{equation*}
G+H = g \mu + (g\sigma + a)W_1 + \frac{b}{2}|W_1|^2.
\end{equation*}
It is straightforward to have
\begin{equation*}
\begin{split}
 v(t,w) =& g \mu + \ (g \sigma + a )w + \frac{b}{2} |w|^2 
-\frac{c\gamma}{2} \frac{(g\sigma + a + bw)^2(1-t)}{c+\gamma + c\gamma b(1-t)} \\ &+ \frac{c+\gamma}{2c \gamma}\log \left(1 + \frac{c\gamma}{c + \gamma} b(1-t) \right), \\
p(t,w,y) = & \ (g-y)\mu + (g-y)\sigma w - \frac{\gamma}{2}(g-y)^2\sigma^2(1-t).
\end{split}
\end{equation*}
It follows then that
\begin{equation*}
\frac{\partial p}{\partial w}(t,w,y) = (g-y)\sigma, \ \ 
\frac{\partial v}{\partial w} =   (g\sigma + a + bw )\frac{c+\gamma}{c+\gamma + c\gamma b (1-t)}.
\end{equation*}
The completeness condition is therefore satisfied and we have
\begin{equation*}
 Y^\ast_t = g - \frac{g\sigma + a + bW_t}{\sigma}\frac{c}{c+\gamma + c\gamma b (1-t)}.
\end{equation*}
The efficient price, the EIPU and the efficient price convexity are  given by
\begin{equation*}
\begin{split}
& P_t(-Y^\ast_t,y) = y(\mu + \sigma W_t) + \frac{\gamma}{2}\sigma^2(1-t)\left\{
(g-Y^\ast_t-y)^2 - (g-Y^\ast_t)^2
\right\},\\
&  S^\ast_t = \mu + \sigma W_t - \gamma \sigma^2(1-t)(g-Y^\ast_t) = \mu + \sigma \frac{(c+\gamma) W_t - (g\sigma + a)c \gamma (1-t)}{c+\gamma + c\gamma b (1-t)},\\
& C_t = \gamma \sigma^2 (1-t).
\end{split}
\end{equation*}
The volatility of $S^\ast$ is therefore given by
\begin{equation*}
 \frac{\mathrm{d}}{\mathrm{d}t} \langle S^\ast \rangle_t = \frac{\sigma^2(c+\gamma)^2}
{(c+\gamma + c\gamma b(1-t))^2}.
\end{equation*}
The volatility monotonically depends on $b$, the convexity of $-H$ as a function of $S$.
In case $b$ is negative, or equivalently,  if the RLD has to hedge an European payoff $-H$ which is a convex function of $S$, then the volatility is larger in case $\gamma > 0$
than in the risk-neutral 
case $\gamma = 0$.
This means that economic signals are amplified in a market where hedgers of convex payoffs are dominant.

\subsection{Ride  a shock wave}
\begin{figure*}
\begin{center}
\includegraphics[width=12.5cm]{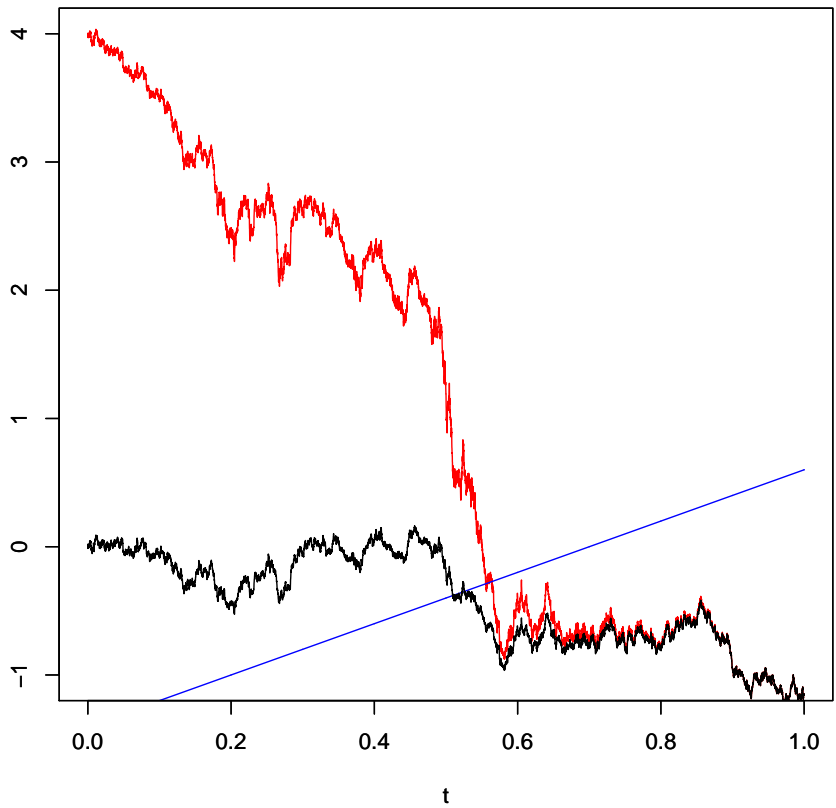}
\caption{A crash of the EIPU (red line) when the factor process $-W$ (black line) meets a shock wave (blue line for the steepest position $-w_c-a(1-t)$): $\mu = 0$, $\sigma = 1$, $a = 2$ and  $w_c=-0.6$.} \label{fig1}
\end{center}
\end{figure*}

Each of the PDEs in (\ref{KPZ}) is a KPZ equation and by differentiating in $w$, we obtain
(backward) Navier-Stokes equations for irrotational flow
\begin{equation*}
 \begin{split}
  &\frac{\partial u}{\partial t} + \frac{1}{2} \Delta u 
= \frac{1}{2}\frac{c\gamma}{c+\gamma} \nabla |u|^2, \ \ u(1,w) =  \nabla (g+h) (w), \\
 & 
\frac{\partial q}{\partial t} + \frac{1}{2} \Delta q
= \frac{1}{2}\gamma \nabla |q|^2, \ \ q(1,w) =  \nabla (g-ys) (w)
 \end{split}
\end{equation*}
for $u = \nabla v$ and $q = \nabla p$
when $s$, $g$ and $h$ are differentiable. In case $k=1$, they are Burgers equations
\begin{equation} \label{Burg}
 \begin{split}
    &\frac{\partial u}{\partial t} + \frac{1}{2} \frac{\partial^2 u}{\partial w^2} 
= \frac{c\gamma}{c+\gamma} u \frac{\partial u}{\partial w}, \ \ u(1,w) =  \frac{\partial}{\partial w}  (g+h) (w), \\
 & 
\frac{\partial q}{\partial t} + \frac{1}{2} \frac{\partial^2 q}{\partial w^2}
= \gamma  q \frac{\partial q}{\partial w}, \ \ q(1,w) =  \frac{\partial}{\partial w} (g-ys) (w).
 \end{split}
\end{equation}
A nontrivial explicit solution for the Burgers equation is known; it is easy to see
\begin{equation*}
 u(t,w) = 1 -\tanh (a (w-w_c) - a^2(1-t)), \ \ a = \frac{c\gamma}{c+\gamma},
\end{equation*}
is a solution with
\begin{equation*}
(g+h)(w) = w - \frac{1}{a} \log(\cosh(a(w-w_c))) + b
\end{equation*}
where $w_c$ and $b$ can be arbitrary constants.
The solution is unique; see Hopf~\cite{Hopf}.
Remark that if $b= - w_c$, then
\begin{equation*}
-(g+h)(w) \to 2(w_c-w)_+
\end{equation*}
as $a\to \infty$. Therefore in case $g = 0$ and $S = \mu - \sigma W_1$ with constants $\mu, \sigma > 0 $, 
the problem of RLD is to hedge a payoff which is close to a call option payoff:
\begin{equation*}
 -H \approx \frac{2}{\sigma} (S-(\mu-\sigma w_c))_+.
\end{equation*}
The completeness condition is satisfied and
\begin{equation*}
 Y^\ast_t = \frac{1}{\sigma}\frac{c}{c+\gamma} \frac{\partial v}{\partial w}(t,W_t)
=  \frac{1}{\sigma}\frac{c}{c+\gamma} (1-\tanh(a(W_t-w_c)-a^2(1-t))).
\end{equation*}
It follows then that
\begin{equation*}
 S^\ast_t = \mu -\sigma W_t + \sigma(1-t)
a(1-\tanh(a(W_t-w_c)-a^2(1-t))).
\end{equation*}
In case that $a = c\gamma/(c+\gamma)$ is large, the function $u$ has a steep slope around $w -a(1-t) = w_c$.
This means that when $W_t - a(1-t)$ approaches to $w_c$ to cross it from below,
the EIPU $S^\ast$ exhibits a drastic drop. See Figure~\ref{fig1} for a sample path.
This can be interpreted as an asset price crash invoked by a propagated endowment shock. \\

The Burgers equation models fluid dynamics and is known to form a shock wave.
More precisely, an inviscid Burgers equation can form a shock wave with a singularity. 
A viscous Burgers equation like (\ref{Burg})
does not form a singularity but can form a steep shape because as $a \to \infty$,
the equation tends to be inviscid.
As in the above explicit case, the EIPU will crash when the factor process $W$ meets a shock wave.
The shock wave does not have a singularity but has a steep slope when $a$ is large, or equivalently, when both the RLS and RLD are strongly risk averse. 
A more detailed analysis on this shock wave model would be worth to be done; it remains for future research.

\section{Dynamic Programming principle}
\subsection{Incomplete markets}
Here we show a version of dynamic programming principle holds in our problem.
Let $D_n = \{0,1/n,\dots, (n-1)/n\}$ with an integer $n > 0$ and 
denote by $\mathcal{S}_A^n$ the set of strategies $Y \in \mathcal{S}_A$ such that
 $\Delta Y_t \neq 0$ only if $t \in D_n$.
In this section, we consider the maximization of $U_0(H + \mathcal{I}(Y))$ subject to
$Y \in \mathcal{S}_A^n$.
As the following lemma shows,
this problem coincides with the original one if 
the filtration $\{\mathcal{F}_t\}$ is discrete in the sense that
$\mathcal{F}_t = \mathcal{F}_{[nt]/n}$ for all $t \in [0,1)$.
This means that information is updated only at a discrete time set.

\begin{lem} \label{lem1}
If $\mathcal{F}_t = \mathcal{F}_{[nt]/n}$ for all $t \in [0,1)$, then
\begin{equation*}
 \sup_{Y \in \mathcal{S}_A}U_0(H + \mathcal{I}(Y))) 
=  \sup_{Y \in \mathcal{S}_A^n}U_0(H +\mathcal{I}(Y)).
\end{equation*} 
\end{lem}
{\it Proof: }
For $t \in [j/n, (j+1)/n)$, putting $\tau = j/n$,
\begin{equation*}
\begin{split}
P_\tau &(-Y_\tau, \eta) + 
P_t(-Y_{\tau} - \eta, y) \\
= &\Pi_\tau(G_m-Y_\tau S)- \Pi_\tau(G_m-(Y_\tau+\eta)) \\ &+ 
\Pi_t(G_m - (Y_{\tau} + \eta)S)-
\Pi_t(G_m - (Y_{\tau} + \eta + y)S)
\\ = &
\Pi_\tau(G_m - Y_{\tau} S)-
\Pi_\tau(G_m - (Y_{\tau} + \eta + y)S)
\\ = & P_\tau(-Y_\tau, \eta + y),
\end{split}
\end{equation*}
which means that buying $\eta$ units at time $\tau$ and $y$ units at
time $t$ is equivalent to buying $\eta + y$ units at time $\tau$.
\hfill////

\begin{prop}\label{prop1}
Define $\{V^n_t\}$ recursively by
\begin{equation*}
V^n_t(x,z)  = \begin{cases}
H + x - zS  & \text{ if } t \geq 1,\\
\sup_{y \in A+z}
U_{[nt]/n}(V^n_{([nt]+1)/n}(x - P_{[nt]/n}(z,y),
-z+y)) & \text{ otherwise}
\end{cases}
\end{equation*}
for $(x,z) \in \mathbb{R} \times (-A)$.
If $A$ is compact,  then
\begin{equation*}
  V^n_0(0,0) = \max_{Y \in \mathcal{S}^n_A}U_0(H + \mathcal{I}(A))
\end{equation*}
\end{prop}
{\it Proof: }
First note that
$V^n(x,z)$ is constant on $[j/n,(j+1))$ for each $(x,z)$ and $j=0,1,\dots n-1$.
Since
\begin{equation*}
U_0(F) = U_0 \circ U_{1/n} \cdots \circ U_{(n-1)/n}(F)
\end{equation*}
for any random variable $F$ and 
\begin{equation*}
\begin{split}
&U_{(j-1)/n}(V^n_{j/n}(-\sum_{0\leq t < j/n}P_t(-Y_t,\Delta Y_t),
 Y_{j/n})) \\
&\leq V^n_{(j-1)/n}(-\sum_{0\leq t < (j-1)/n}P_t(-Y_t,\Delta Y_t),Y_{(j-1)/n})
\end{split}
\end{equation*}
for any $Y \in \mathcal{S}_A^n$ and $j=1,2,\dots n$, we obtain
\begin{equation*}
  V^n_0(0,0) \geq \sup_{Y \in \mathcal{S}^n_A}U_0(H + Y_1S - \sum_{0 \leq t
   < 1} P_t(-Y_t, \Delta Y_t)).
\end{equation*}
On the other hand, since $A$ is compact,
 there exists a
sequence  $\{Y^\ast_j\}$ adapted to
$\{\mathcal{F}_{j/n}\}$ such that 
\begin{equation*}
  V^n_0(0,0) = U_0(H + Y_1S - \sum_{0 \leq t
   < 1} P_t(-Y_t, \Delta Y_t))
\end{equation*}
with
\begin{equation*}
Y_t = \sum_{ j\leq [nt]} Y^\ast_j
\end{equation*}
by the measurable selection theorem (see Parthasarathy~\cite{Par}). \hfill////

\begin{lem}
Let $F$ be a random variable and $t \in D_n$. Define  a sup-convolution operator $\Psi_t$ as
\begin{equation*}
 \Psi_t(F) = 
\sup\{ U_t(F-L) + \Pi_t(L); L \in \{\Pi_{t+1/n}(G-yS); y \in A\}\}.
\end{equation*}
Then, for any
 $\mathcal{F}_t$-measurable random vector $Z=(Z^1,\dots,Z^d)$,
 \begin{equation*} 
\begin{split}
  \sup_{y \in A + Z} &U_t(F-\Pi_{t+1/n}(G + (Z-y)S+P_t(Z,y)))
\\ &= \Psi_t(F) - \Pi_t(G + ZS).
\end{split}
 \end{equation*}
\end{lem}
{\it Proof: } 
Since
\begin{equation*}
\begin{split}
&  \Pi_{t+1/n}(G + (Z-y)S + P_t(Z,y)) \\ 
&= \Pi_{t + 1/n}(G + (Z-y)S) +
 P_t(Z,y) \\ &=
\Pi_{t + 1/n}(G + (Z-y)S) +
 \Pi_t(G + ZS) - \Pi_t(G + (Z-y)S)
\end{split}
\end{equation*}
and $\Pi_t(G + (Z-y)S) = \Pi_t(\Pi_{t+1/n}(G + (Z-y)S))$,
\begin{equation*}
\begin{split}
&U_t(F-\Pi_{t+1/n}(G + (Z-y)S+P_t(Z,y)))
+ \Pi_t(G + ZS) \\
&= U_t(F-\Pi_{t+1/n}(G + (Z-y)S)) +
\Pi_t(\Pi_{t+1/n}(G + (Z-y)S)).
\end{split}
\end{equation*}
Taking $\sup$ in $y$, we obtain the representation. \hfill////

\begin{prop} \label{repsup}
\begin{equation*}
V^n_0(0,0) +  \Pi_0(G)
= \Psi_{0} \circ \Psi_{1/n} \circ \cdots \Psi_{(n-1)/n}(G+H).
\end{equation*}
\end{prop}
{\it Proof: } Note that
\begin{equation*}
V^n_1(x,z) = H + x -z S  = x + G+H - \Pi_1(G+zS).
\end{equation*}
Then, applying the previous lemma with $F = G+H$,
\begin{equation*}
\begin{split}
V^n_{(n-1)/n}(x,z) =& x + \sup_{y \in A+z}U_{(n-1)/n}(F
- \Pi_1(G+(z-y)S  + P_{(n-1)/n}(z,y))
) \\
=& x + \Psi_{(n-1)/n}(F) - \Pi_{(n-1)/n}(G +zS).
\end{split}
\end{equation*}
Repeat this to get the representation. \hfill////\\

\subsection{No rebalancing solution}
As a direct application of the above results, let us consider a special case that there exists $y^\ast \in A$ such that
\begin{equation*}
G-y^\ast S = \frac{c}{c+\gamma}(G+H), \ \ 
H + y^\ast S = \frac{\gamma}{c+\gamma}(G+H).
\end{equation*}
This holds, for example, both $G$ and $H$ are proportional to $S$ and $A$ is sufficiently large. 
Assume  $G$, $H$ and $S$ are bounded and $A$ is compact.
Applying  the preceding results, we have
\begin{equation*}
\max_{Y \in \mathcal{S}^n_A}U_0(H+\mathcal{I}(Y)) = \Psi_0 \circ \cdots \Psi_{(n-1)/n}(G+H) - \Pi_0(G).
\end{equation*}

Let us observe that the supremum is attained by
\begin{equation*}
Y_t = y^\ast\ \ t \in (0,1], \ \ Y_0 = 0.
\end{equation*}
This means that the RLD buys $y^\ast$ units of the securities at time $0$ and holds them until time $1$.
The supremum in 
\begin{equation*}
\Psi_{(n-1)/n}(G+H) = 
\sup_{y \in A}\{ U_{(n-1)/n}(H+yS) +
\Pi_{(n-1)/n}(G-yS) \}
\end{equation*}
is attained by $y^\ast \in A$ because
\begin{equation*}
 y \mapsto U_{(n-1)/n}(H+yS) +
\Pi_{(n-1)/n}(G-yS)
\end{equation*}
is concave and
\begin{equation*}
\frac{\partial}{\partial y^i}
\{ U_{(n-1)/n}(H+yS) +
\Pi_{(n-1)/n}((G-yS) \}\bigg|_{y = y^\ast} = 0
\end{equation*}
for all $i=1,\dots,d$.
Further we have
\begin{equation*}
\begin{split}
\Psi_{(n-1)/n}(G+H) =& 
- \frac{c + \gamma}{c\gamma} \log
\mathbb{E}\left[
\exp\left\{ - \frac{c\gamma}{c + \gamma}(G+H)\right\} \bigg|
\mathcal{F}_{(n-1)/n}
\right] \\
= & \frac{c+\gamma}{c} \Pi_{(n-1)/n}\left(\frac{c}{c+\gamma}(G+H)\right).
\end{split}
\end{equation*}
Then we go to the next step. We have
\begin{equation*}
\begin{split}
&\Psi_{(n-2)/n}(\Psi_{(n-1)/n}(G+H)) \\&= 
\sup_{y \in A} \{
U_{(n-2)/n}(
 F - \Pi_{(n-1)/n}(G-yS)) + \Pi_{(n-2)/n}(G-yS)
\} 
\\ & \leq \sup_{H \in \mathcal{L} } \{
U_{(n-2)/2}(F - \Pi^\ast - L) + \Pi_{(n-2)/2}(\Pi^\ast + L)
\},
\end{split}
\end{equation*}
where
\begin{equation*}
\begin{split}
&F = \frac{c+\gamma}{c} \Pi_{(n-1)/n}\left(\frac{c}{c+\gamma}(G+H)\right)
\\
&\Pi^\ast =  \Pi_{(n-1)/n}(G-y^\ast S) = \Pi_{(n-1)/n}\left(\frac{c}{c+\gamma}(G+H)\right) = \frac{c}{c+\gamma}F
\end{split}
\end{equation*}
and $\mathcal{L}$ is the linear space spanned by 
$\Pi_{(n-1)/n}(G-yS)$, $y\in A$.
The function
\begin{equation*}
 \epsilon \mapsto U_{(n-2)/2}(F - \Pi^\ast -\epsilon L) + \Pi_{(n-2)/2}(\Pi^\ast + \epsilon L)
\end{equation*}
is concave for each $L \in \mathcal{L}$ and 
its derivative in $\epsilon$ at $\epsilon = 0$ vanishes due to that
 $c(F- \Pi^\ast) = \gamma \Pi^\ast$.
Therefore
\begin{equation*}
\begin{split}
&\sup_{y \in A} \{
U_{(n-2)/n}(
 F - \Pi_{(n-1)/n}(G-yS)) + \Pi_{(n-2)/n}(G-yS)
\} \\& \leq 
U_{(n-2)/2}(F - \Pi^\ast) + \Pi_{(n-2)/2}(\Pi^\ast) 
\end{split}
\end{equation*}
and the upper bound is attained when $y = y^\ast$.
Consequently we get
\begin{equation*}
\Psi_{(n-2)/n}(\Psi_{(n-1)/n}(G+H)) =
- \frac{c + \gamma}{c\gamma} \log
\mathbb{E}\left[
\exp\left\{ - \frac{c\gamma}{c + \gamma}(G+H)\right\} \bigg|
\mathcal{F}_{(n-2)/n}
\right]
\end{equation*}
and apparently this argument can be repeated to conclude that the optimal
strategy to hold $y^\ast$ units from the beginning.\\

Consequently, the optimal strategy is $Y^\ast_t = y^\ast$ for $t > 0$, which implies
\begin{equation*}
\begin{split}
 \max_{Y \in \mathcal{S}_A^n}U_0(H + \mathcal{I}(Y)) &= U_0( H + y^\ast S - P_0(0,y^\ast)) \\
&= U_0(H+y^\ast S ) + \Pi_0(G-y^\ast S) - \Pi_0(G) \\
&= U^\ast_0(G+H) - \Pi_0(G).
\end{split}
\end{equation*}
This extends (\ref{alloc}) and means that the optimal risk allocation is achieved.
Denote by $Q$ the probability measure defined by
\begin{equation*}
\frac{ \mathrm{d}Q }{\mathrm{d}\mathbb{P}} = \frac{\Gamma}{\mathbb{E}[\Gamma]}, \ \ 
\Gamma = \exp\left\{
-\frac{c\gamma}{c+\gamma}(G+H)
\right\}.
\end{equation*}
Then,
\begin{equation}\label{EMM}
 S^\ast_t = \frac{\mathbb{E}[S\exp\{-\gamma(G-y^\ast S)\}|\mathcal{F}_t]}
{\mathbb{E}[\exp\{-\gamma(G-y^\ast S)\}|\mathcal{F}_t]} = \mathbb{E}^Q[S | \mathcal{F}_t].
\end{equation}
Further, the efficient price convexity coincides with the conditional covariance matrix of $S$ under $Q$:
\begin{equation*}
 C^{ij}_t = \mathbb{E}^Q[S^iS^j |\mathcal{F}_t] -
\mathbb{E}^Q[S^i|\mathcal{F}_t] \mathbb{E}^Q[S^j |\mathcal{F}_t].
\end{equation*}
We may have these simple expressions for $S^\ast$ and $C^{ij}$ because the optimal strategy is of buy-and-hold type.\\

The boundedness condition on $S$, $G$ and $H$ can be relaxed.
Due to (\ref{EMM}), standard models of mathematical finance are then supported in this framework.
For example, if $d=1$ and
\begin{equation*}
S = \mu + \sigma W_1, \ \ H = \alpha S, \ \ G = \beta S
\end{equation*}
with a standard Brownian motion $W$ and 
 constants $\alpha, \beta, \mu  \in \mathbb{R}$ and $ \sigma > 0$, then
 we
obtain the Bachelier model
\begin{equation*}
\mathrm{d}S^\ast_t = \frac{c\gamma }{c + \gamma}(\alpha + \beta)\sigma^2 \mathrm{d}t +
 \sigma \mathrm{d}W_t, \ \  
S^\ast_0 =  \mu - \frac{c}{c + \gamma}(\alpha + \beta)\sigma^2.
\end{equation*}
If $d=1$ and
\begin{equation*}
S = \zeta e^{\sigma W_1}, \ \ G  = \alpha \sigma W_1 - H,  \ \ 
H = \frac{\gamma}{c + \gamma} \alpha \sigma W_1- 
\frac{\mu}{c+\gamma}S
\end{equation*}
with constants $\alpha, \mu \in \mathbb{R}$ and $\sigma, \zeta > 0$, then we
have $Y^\ast_t = \mu/(c + \gamma)$ and obtain
the Black-Scholes model
\begin{equation*}
\mathrm{d}S^\ast_t = \frac{c \gamma }{c + \gamma } \alpha\sigma^2 S^\ast_t
 \mathrm{d}t + \sigma S^\ast_t \mathrm{d}W_t, \ \ 
S^\ast_0 = \zeta \exp\left\{\sigma^2\left(\frac{1}{2}-
				\frac{c\gamma}{c+\gamma}\alpha\right)\right\}.
\end{equation*}

\subsection{Complete Markov models revisited}
Here we reconsider the framework of Section~4.4 in terms of the dynamic programming principle.
Our aim here is to recover Theorem~\ref{thm:opt} by letting $n\to \infty$.
The filtration $\{\mathcal{F}_t\}$ is generated by a $k$-dimensional $\{\mathcal{F}_t\}$-standard Brownian motion $W$.
 We suppose that $A$ is a compact set and
\begin{equation*}
S = s(W_1), \ \ G = g(W_1), \ \   H=h(W_1)
\end{equation*}
 with functions $s : \mathbb{R}^k \to \mathbb{R}^d$, $g:\mathbb{R}^k \to \mathbb{R}$ and $h:\mathbb{R}^{k} \to \mathbb{R}$.
Here we assume in addition the boundedness and 
the uniform continuity for the derivatives
\begin{equation*}
\frac{\partial v}{\partial t}, \ \ 
\frac{\partial v}{\partial w_i}, \ \ 
\frac{\partial^2 v}{\partial w_i \partial w_j}, \ \
\frac{\partial p}{\partial t}, \ \ 
\frac{\partial p}{\partial w_i}, \ \  
\frac{\partial^2 p}{\partial w_i \partial w_j}, \ \ i,j=1,\dots,k.
\end{equation*}
for $p$ and $v$ defined in Section~4.4.
\begin{lem}
Under the completeness condition (\ref{complete:markov}), 
\begin{equation*}
\lim_{n \to \infty} 
n \max_{t \in D_n}
\|v(t,W_t)  - \Psi_t(v(t+1/n),W_{t+1/n})\|_{\infty} = 0.
\end{equation*}
\end{lem}
{\it Proof: }
By (\ref{markov}),
\begin{equation*}
p(t,W_t,y) = \Pi_t(p(t+1/n,W_{t+1/n},y)).
\end{equation*}
By Taylor's theorem,
\begin{equation*}
\begin{split}
p&(t,w,y)\\ =&
- \frac{1}{\gamma} \log
\mathbb{E}\left[\exp\left(-\gamma 
p(t+1/n,W_{t + 1/n},y)
\right) \big|W_t=w\right]
\\ =& p(t+1/n,w,y)  \\ & -\frac{1}{\gamma} \log
\biggl\{
1 
-\frac{\gamma}{2n} \Delta p(t+1/n,x,y)  
 + \frac{\gamma^2}{2n}\left| \nabla p (t+1/n,w,y)\right|^2 + o(1/n)
\biggr\}
\\
=&
p(t+1/n,w,y) 
+ \frac{1}{2n} \Delta p(t+1/n,w,y) 
- \frac{\gamma}{2n}\left| \nabla p (t+1/n,w,y)\right|^2 + o(1/n).
\end{split}
\end{equation*}
Here and hereafter,
the estimate $o(1/n)$ is uniform in $(t,w,y)$.
Similarly, 
\begin{equation*}
\begin{split}
-& \frac{1}{c} \log
\mathbb{E}\left[\exp\left(-c \left(
v(t+1/n,W_{t + 1/n}) - p(t+1/n,W_{t+1/n},y) \right)
\right) \big|W_t=w\right] \\ 
= & 
v(t+1/n,w)- p(t+1/n,w,y)  + \frac{1}{2n} 
\left(\Delta v(t+1/n,w) -
\Delta p(t+1/n,w,y) 
\right) \\ &- \frac{c}{2n}\left|
\nabla v(t+1/n,w) - \nabla p(t+1/n,w,y)\right|^2 + o(1/n).
\end{split}
\end{equation*}
Therefore,
\begin{equation*}
\begin{split}
\Psi_t&(v(t+1/n,W_{t+1/n})) \\ =&  \sup_{y \in A} \biggl\{
v(t+1/n,W_t) + \frac{1}{2n} 
\Delta v(t+1/n,W_t) - \frac{c}{2n}
\left|\nabla v (t+1/n,W_t)\right|^2  \\& +
\frac{c}{n} 
(\nabla v(t+1/n,W_t), \nabla p(t+1/n,W_t,y)) \\& -
\frac{c+ \gamma}{2n}\left|
\nabla p(t+1/n,W_t,y)\right|^2
 + o(1/n) \biggr\} \\
=&  
v(t,W_t) - 
\inf_{y \in A} \biggl\{
\frac{c + \gamma}{2n} 
\left|
\nabla p(t+1/n,W_t,y)
-\frac{c}{c+\gamma}\nabla v(t+1/n,W_t)
\right|^2
+ o(1/n) \biggr\}.
\end{split}
\end{equation*}
Then the result follows from the completeness condition. \hfill////

\begin{thm} \label{thm1}
Under the completeness condition (\ref{complete:markov}), 
\begin{equation*}
\lim_{n \to \infty } V^n_0(0,0) = v(0,W_0) - p(0,W_0,0) =
U^\ast_0(G+H) - \Pi_0(G).
\end{equation*}
 \end{thm}
{\it Proof: }
Put
\begin{equation*}
r_n = \sup_{t \in D_n}\|v(t,W_t)  - \Psi_t(v(t+1/n),W_{t+1/n})\|_{\infty}.
\end{equation*}
By the previous lemma, $nr_n \to 0$ as $n \to \infty$.
For $k=0,1,\dots, n-1$, put 
\begin{equation*}
F_k = \Psi_{k/n} \circ
\Psi_{(k+1)/n} \circ \cdots \circ \Psi_{(n-1)/n}(G+H).
\end{equation*}
Then, by the cash-invariance property of $\Psi_{k/n}$,
\begin{equation*}
\begin{split}
\|
F_k - v(k/n,W_{k/n})
\|_{\infty}  \leq &
\|
\Psi_{k/n}(F_{k+1}) -\Psi_{k/n}(v((k+1)/n,W_{(k+1)/n}))
\|_{\infty} \\ & +
\|\Psi_{k/n}(v((k+1)/n),W_{(k+1)/n})  - 
 v(k/n,W_{k/n})
\|_{\infty} \\
\leq & \|F_{k+1} - v((k+1)/n,W_{(k+1)/n})\|_{\infty} + r_n.
\end{split}
\end{equation*}
Then, by Proposition~\ref{repsup},
\begin{equation*}
\begin{split}
&|V^n(0,0) - v(0,X_0) + p(0,X_0,0)|
\\ &=
|F_0 - v(0,X^n_0)| \leq  nr_n \to 0
\end{split}
\end{equation*}
as $n\to \infty$. \hfill////

\end{document}